\documentclass[12pt]{iopart}
\usepackage{epsfig,pslatex,amssymb,iopams}

\def\gm{g_{-}}

\def\gp{g_{+}}
\def\emm{e_{-}}

\def\ep{e_{+}}
\def\lr{\leftrightarrow}
\def\ra{\rangle}

\def\ch{\chi^{(1)}}
\begin{document}
\title{Tunable nonlinearity in atomic response to a bichromatic field}

\author{Ashok
Vudayagiri\footnote{Present address: Raman Research Institute, C. V. Raman
Avenue, Sadashivanagar, Bangalore 560080, India: e-mail: ashok@rri.res.in} and Surya P. Tewari}
\address{School of Physics,
University of Hyderabad, Hyderabad, 560146 India }
\ead{\mailto{ashok@rri.res.in},\mailto{sptsp@uohyd.ernet.in}}

\begin{abstract}
Atomic response to a probe beam can be tailored, by creating coherences between atomic levels with help of
another beam. Changing parameters of the control beam will change the nature of coherences and hence the nature
of atomic response as well. Such change can depend upon intensity of both probe and control beams, in a
nonlinear fashion. We present a situation where this nonlinearity in dependence can be precisely controlled, as
to obtain different variations as desired. We also present a detailed analysis of how this nonlinear dependency
arises and show that this is an interesting effect of several Coherent Population Trap(CPT) states that exist
and a competition among them to trap atomic population in. 
\end{abstract}
\pacs{42.50.Gy, 42.50.Hz, 42.50.Nn}
\submitto{\JPB}

Response of a material medium to an electromagnetic field usually exhibits a linear dependence on the
intensity, at low intensities. When the intensity is sufficiently high, intensity dependant components of
refractive index and absorption become dominant and the response of the material starts becoming increasingly
nonlinear. While this is usually true for bulk medium, dilute atomic gases show a nonlinearity even at lower
intensities. This is due to the fact that a near resonant laser prepares the
atom in a coherent superposition of its energy levels. The response of an atom in such a superposition is not
merely a sum of responses from the atom in its constituent states and can be  dramatically different.
 Thus, by creating a superposition state, a control laser can alter the behaviour of
the atom towards a probe laser propagating through such a medium. Such coherence induced control of an atomic
response can also be interpreted as a coupling between photons of the control beam and those of the probe
beam, giving us the same picture as in traditional nonlinear optics \cite{harris},\cite{petrosyan}. Although a
nonlinear response at higher intensities is also observed, due to anharmonic interaction of the atom with the
laser, the coherence induced nonlinearity is much more efficient since it happens at much lower intensities.
This phenomena has found several applications such as achieving giant nonlinearities \cite{kerr}, efficient
generation of of Vacuum UltraViolet (VUV) using third harmonic\cite{spt}, suppression of nonlinear  responses
\cite{maneesh}, light induced focusing \cite{dunn} and many more. A fast variation of atomic response between
a complete transparency to a complete opacity can also be achieved which would effectively make an optical
switch using the medium \cite{switch}. 

A recent study \cite{strekalov} experimentally investigates several $\Lambda$-like systems for dependency of
atomic response on probe intensity and shows existence of a threshold intensity where linearity breaks down. On
the other hand, we present a slightly different configuration in this paper which exhibits an almost completely
nonlinear response. Further, this nonlinearity is precisely controllable, by manipulating the intensity and
frequency of the pump beam. Our configuration consists of two hyperfine states with quantum
number F=1 and its Zeeman manifolds, totally amounting to six levels. Such a system is often encountered in
atoms such as $^{87}$Rb or Sodium, which is used in many atom-laser experiments. In other words, this is a
familiar system which exhibits a sufficiently rich nonlinear response under appropriate conditions, as
explained in this paper. We also present in detail the algorithm used to obtain numerical solutions which
describe the dynamics of this system. Due to involvement of two lasers of slightly differing frequencies, the
most used method of transformation into a rotational frame will not eliminate explicit time-dependencies in the
interaction Hamiltonian and hence we resort to a continued fraction approach to solve them. This method, being
non-perturbative in approach is valid for any intensities of pump and probe beams. Thus, we are not confined to
regimes of strong pump and weak probe alone, and are able to continuously investigate the behaviour of the atom
as probe beam's intensity is increased from almost zero value to a value larger than that of the pump beam.
Augmenting the study with an analysis using Semi-classical Dressed states, we are able to show that the
complex response shown by the atom is due to existence of many Coherent Population Trap (CPT) states which
compete with each other for atomic occupation.

CPT states are formed when superposition of atoms are prepared in such a superposition of its bare states, that
the probability of transitions from each of its constituent states to a common external state cancel each
other. Thus, the atom can not absorb a photon to make a transition to the common external state is thus
trapped \cite{arimondo}. However, there may exist other states to which transitions not canceled, and the CPT
state can be depopulated by using the appropriate laser. Our configuration, due to existence of many degenerate
levels, leads to formation of more than one CPT state. As the parameters of the lasers, such as frequency and
intensity are scanned, the atoms are transferred from one CPT state to other. The rates of optically pumping
the population to a particular CPT state or out of it depends upon the nature of the state. Therefore, in the
intermediate states, a competition between different CPT states manifests, as to trapping the population. This
phenomena is observed as a variation in probe absorption and dispersion, which varies nonlinearly with
variation of intensities of the beam, as shown in this paper. Furthermore, the nature of nonlinear response
shown to one of the beams can be modified by varying the detuning or intensity of the other beam involved.
Thus, a precise control over the nonlinearity is possible, which  may have several applications in information
processing using nonlinear properties of the medium.  

This communication is divided into three parts. We describe the atomic configuration in detail in the following
section, and also describe the method we have adopted to solve the density matrix equations. In the subsequent
section, we present the results obtained under different conditions of pump and probe intensities. From this
numerical results, we attempt an insight into the phenomena that occurs. We obtain the semiclassical dressed
states in section IV and substantiate our conclusion of competition between different CPT states. 

\section{The configuration}
The atomic level configuration consists of six
levels as shown in figure 1. The levels labeled $|g_0\ra,~|g_{+}\ra,~|g_{-}\ra$ form the triplet
of the ground state F$_g$=1 state with magnetic quantum numbers m$_{\rm F}=0,~\pm 1$
respectively. Those, that are labeled  $|e_0\ra,~|e_{+}\ra$ and $|e_{-}\ra$ are similarly the Zeeman sublevels 
of the excited state F$_e=1$, with their magnetic quantum numbers
m$_{\rm F}=0,~1$ and $-1$. For sake of convenience the levels  F$_g$=1 and F$_e=1$ could be
hyperfine levels of the ground state 5S$_{1/2}$ and of the first excited state
5P$_{1/2}$, respectively of $^{87}Rb$. Any two hyperfine states with F=1, of any other alkali atom,
such as Sodium for instance, will give identical results. 

Two beams, plane polarized in orthogonal directions are incident
on this system. For the probe beam, we use the one which is plane polarized along the z-axis of quantisation.
This beam couples the transitions $|{\rm g}_- \ra
\leftrightarrow~|{\rm e}_-\ra$ and $|{\rm g}_+\ra \leftrightarrow~|{\rm e}_+\ra$. The dipole transition between $|{\rm
g}_0 \ra \leftrightarrow~|{\rm e}_0\ra$ is forbidden due to vanishing Clebsch-Gordon coefficients.
The other beam is plane polarized in xy plane, thus forming a combination of two circularly polarized lights
with helicities $\sigma_+$ and $\sigma_-$. This beam therefore couples transitions $|{\rm g}_\pm \ra \leftrightarrow {\rm
|e}_0\ra$ and $|{\rm g}_0 \ra\leftrightarrow |{\rm e}_\pm \ra$. We refer to this as the pump beam. We then
analyse the response of the medium to the probe beam, as modified by the pump beam. 

In the semi-classical approach, the pump and the probe beam are represented by their electromagnetic fields
$E_l={\cal E}_l \exp(i \omega_l t-k_l.z)+c.c.$ and $E_p={\cal E}_p \exp(i \omega_p t-k_p.y)+c.c.$
respectively, where ${\cal E}_l$ is the slowly varying envelop of the pump field and ${\cal E}_p$, that of the
probe field. $\omega_l$ and $\omega_p$ are their respective angular frequencies. The pump beam is assumed to
propagate in z direction,  while probe beam can propagate either in x or y directions to give the required
polarization. We assume it to be y direction for the sake of convenience. For rest of this paper, we refer
these two beams by their respective Rabi frequencies $\Omega_{l,p}=d_{ij}.{\cal E}_{l,p}$ where
$i=|e_{0,\pm}\ra$ and $j=|g_{0,\pm}\ra$.

A closely resembling configuration has been analyzed and reported by Petrosyan et. al., \cite{petrosyan}
showing an efficient two-photon process occurs in the media, with one photon from the pump beam and another
from the probe.  The authors present study of
propagation effects in a situation where the $|g_\pm\ra$ and $|e_\pm \ra$ levels are Zeeman shifted. Our
analysis differs from theirs in three aspects. Firstly we look at the steady state phenomena, which can model
most experiments involving cw lasers as well as very long pulses. Secondly we analyze a nearly-degenerate
system with no Zeeman shifts involved, which, as we show, exhibits a wide variety of non-linear phenomena.
Thirdly, we derive a general method for obtaining numerical solutions of such systems with no restrictions on
intensities of pump and probe beams. The method obtains results to all orders of magnitude in pump and probe
beams and hence can be adopted even when probe and pump beams are of comparable intensities, or even when probe
is stronger than the pump. We present the procedure involved to obtain  numerical solutions below.

The Hamiltonian governing the system is given by 
\begin{equation}
{\cal H}={\cal H}_0+{\cal H}_{int},
\label{hamiltonian}
\end{equation}
where,

\begin{eqnarray}
{\cal H}_0&=&  \omega_{e}|e_+\rangle \langle e_+| +   \omega_{e}
|e_- \rangle \langle e_-| +  \omega_{e} |e_0 \rangle \langle e_0| \cr
&+&   \omega_{g}|g_+\rangle \langle g_+| +   \omega_{g}
|g_- \rangle \langle g_-| +  \omega_{g} |g_0 \rangle \langle g_0|,
\label{h0}
\end{eqnarray}

\begin{eqnarray}
{\cal H}_{\rm int}=&-&  \Omega_- e^{-i \omega_l t+ i k_l z}
(|e_0\rangle \langle g_+| + |e_-\rangle \langle g_0|) \cr
&-&  \Omega_+ e^{-i \omega_l t+ i k_l z}
(|e_0\rangle \langle g_-| + |e_+\rangle \langle g_0|) \cr
&-&  \Omega_p e^{-i \omega_p t+ i k_p y}
(|e_+\rangle \langle g_+| - |e_-\rangle \langle g_-|) + H. C..
\label{hint}
\end{eqnarray}

$\omega_{g,e}$ are the eigen-energies of the states $|g_{0,\pm} \rangle$
and $|e_{0,\pm}\rangle$ respectively in terms of angular frequencies and
$\hbar=1$. The coupling factors are defined as
\begin{eqnarray*}
\Omega_\pm&=&\Omega_l=d_{ab}{\cal E}_l\\
\Omega_p&=&d_{ab}{\cal E}_p \\
{\rm for}~ a&=&|e_{0,\pm}\rangle~{\rm and}~b=|g_{0,\pm}\rangle
\end{eqnarray*}

Relative sign difference for transition between $|e_+\rangle \leftrightarrow
|g_+\rangle$ and $|e_-\rangle \leftrightarrow |g_-\rangle$ in (\ref{hint}) arises
due to a sign difference between dipole matrix elements such that
$$d_{{e_-}{g_-}}=-d_{{e_+}{g_+}}.$$

The evolution of 6x6 density matrix of the system is governed by the Liouville
equation
\begin{equation}
\partial_t \rho=-i  [{\cal H}, \rho] + {\cal L} \rho\label{liouville}.
\end{equation}

${\cal L} \rho$ denote the terms involving decay terms that originate from both spontaneous emission and
collisional mixing. We assume that collisions do not bring about any significant mixing between $|e\rangle$
and $|g\rangle$ states although collisional transfer within the Zeeman substates of $|e\rangle$ and $|g\rangle$ are significant. That
is, most collisions are of the spin-flip kind. The equations for each term of density matrix $\rho$ are
explicitly written in Appendix. Traditional method of obtaining solutions for the equations involve a
transformation into a rotating frame of reference such that the right hand side of equation (\ref{liouville})
has constant coefficients. However using either of the two frequencies involved, namely that of probe or the
pump beam, have oscillating coefficients. In other words, there exist two rotating frames of reference with
rotational frequencies of $\omega_l$ and $\omega_p$, and choosing either one of them as stationary will result
in the other one rotate at a frequency equal to the difference between the two of them.

To solve such a system, we chose a rotational transformation with the pump frequency
$\omega_l$ such that the transitions connected by pump beam is seen
stationary\footnote{Choosing the system which is resonant with $\omega_p$ would
equally serve.}. These are the transitions $|g_-\ra \lr |e_0\ra \lr |g_+\ra$ and $|e_-\ra \lr |g_0\ra
\lr |e_+\ra$, which are the three levels configurations $\Lambda$ and V, which are
independently studied in the past. The probe beam couples these two subsystems whose
transitions are now investigated.

Accordingly, a rotational transformation is operated such as

\begin{eqnarray*}
\rho_{ij} &\rightarrow& \bar \rho_{ij}\exp(-i\omega_l t);~{\rm for}~i\ne j\\
\rho_{ii} &\rightarrow& \bar \rho_{ii}\\
&&{\rm for}~i,j=1,\ldots,6
\label{rotation}
\end{eqnarray*}

The equations in such a rotated frame become
\begin{equation}
\partial_t \bar \rho=A \bar \rho + e^{i (\omega_l -\omega_p) t}M_+ \bar
\rho+e^{-i (\omega_l -\omega_p) t}M_- \bar \rho + I_0 \label{eq1}.
\end{equation}

$A$ is a 35 x 35 matrix that involves all interaction terms which are due to
$\omega_l$ field alone, and those related to free evolution. These terms  are seen as stationary in the
rotating frame chosen. On the other hand, terms in $M_+$ and $M_-$, which are also 35 x 35 matrices contain
terms coupled to $\Omega_p$ field and thus have terms with $\omega_l-\omega_p$ in the exponent. Terms with a
positive exponent (i.e., rotation with a frequency of $+(\omega_l-\omega_p)$ are grouped in $M_+$ and those
with a rotation in the opposite sense are grouped in $M_-$. These terms are go to zero when $\Omega_p$ is zero.

$I_0$ is the single column  matrix of  dimension (35x1). The column $I_0$ depends on the way the constraint

\begin{equation}
\rho_{ee}+\rho_{gg}=1\label{constr}.
\end{equation}

is used to write (\ref{eq1}). In (\ref{eq1}) the element $\rho_{g_0 g_0}$ has
been eliminated using (\ref{constr}). Explicit values for $A,~M_+,~M_-$ and $I_0$ that are used in our
calculations are given in appendix.

The solution of (\ref{eq1}) is obtained
by assuming
\begin{equation}
\bar \rho=\sum_{n=-\infty}^{+\infty} e^{in(\omega_l-\omega_p)t} \bar
\rho^{(n)}. \label{series}
\end{equation}
Substituting (\ref{series}) in (\ref{eq1}) and equating synchronous terms on both
sides of (\ref{eq1}), one obtains
\begin{equation}
\partial_t \bar \rho^{(n)}=\left[A-i~n(\omega_l-\omega_p)\mathbb{I}\right]
\bar \rho^{(n)}+M_+\bar \rho^{(n-1)}+M_- \bar \rho^{(n+1)} + I_0 \delta_{n0}.
\end{equation}
$\mathbb{I}$ is the 35 x 35 identity matrix.

These equations determine the dynamics of the different subsystems that exist
in our main system. The transfer of populations and coherences from one
subsystem to the other can be determined from here. In this paper we present
results in steady state. The steady state is described as the state that is
created as a result of equilibrium reached due to the continuous exchange
of populations and coherences among different subsystems due to the
incoherent processes like collisions and spontaneous emission
as well as coherent interactions due to the two fields. Such a steady state
is characterized mathematically by setting on left hand side
$\partial_{t} \bar \rho^{(n)}=0$. Using this condition leads to three
term algebraic recursion relations \cite{risken}, which can be solved in the
following way:

We define square matrices $X_n,~X_{-m}$ such that
\begin{eqnarray}
\bar \rho^{(n)}=X_n\bar \rho^{(n-1)};~~~~~~~~~~n>0 \label{eq2}\\
\cr
\bar \rho^{(-m)}=X_{-m}\bar \rho^{(-m+1)};~~~~~~~~~~m>0 \label{eq3}\
\end{eqnarray}
and use these in the recursion relations for different $n$, $m$. One finds
\begin{eqnarray}
X_n=-\frac{M_+}{A-i~n(\omega_l-\omega_p)\mathbb{I}+M_- X_{n+1}} ,\label{eq4}\\
\cr
X_{-m}=-\frac{M_-}{A+i~m(\omega_l-\omega_p)\mathbb{I}+M_+ X_{-m-1}} ,\label{eq5}\
\end{eqnarray}

then $\bar \rho^{(0)}$ is given by

\begin{eqnarray}
\bar \rho^{(0)}=\frac{-I_0}{A+M_+ X_{-1}+ M_- X_{+1}}  .\label{eq6}\
\end{eqnarray}

This completes consistent mathematical procedure used for the numerical study
for variety of cases in the six level  model. The numerical procedure involves
truncating the recursion relation (\ref{eq4}) and (\ref{eq5}) for sufficiently
large values of integers n and m such that the result of (\ref{eq6}) converge to
stable values. These are then used to determine other observables of the system 
using (\ref{eq2}) and (\ref{eq3}) above. This method, which is
non-perturbative in approach, can be used for arbitrary intensities of pump and
probe beams, as long as result of (\ref{eq6}) converges. The methodology adopted to solve this system can be
extended to study several other configurations which involve two or more lasers and leading to time dependent
terms such as those found in this case. We could use the same mathematics to solve a more complicated systems
where the pump laser interacted resonantly with some transitions and off-resonantly with others, which were
also coupled by the probe beam. We got consistent solutions for that system as well.

\section{Results of numerical solutions}
Using the method discussed in previous section, we numerically solve equation (\ref{eq6})
and obtain total susceptibility as 
\begin{equation}
\chi^{(1)}(\omega_p,{\cal E}_p)=\frac{d_{\gp\ep}^{(+1)}-d_{\emm\gm}^{(+1)}}{{\cal E}_p}.
\label{aa}
\end{equation}

Since absorption and refractive index are related to $\chi^{(1)}$ as inverse Beer length
$\alpha=2\pi(\omega/c){\rm Im}\chi^{(1)}$  and $n=1+2\pi ~{\rm Re} \chi^{(1)}$ respectively, we compute real
and imaginary parts of $\ch$ to qualitatively study the atomic response.  We present real and imaginary parts
of $\chi^{(1)}$ as a function of probe Rabi frequency $\Omega_p$, for different values of pump Rabi frequency
$\Omega_l$. We notice a pronounced nonlinear variation of both absorption and dispersion of the probe beam. For
the graphs shown in figure 2, the pump beam is assumed to be resonant with the $|g\ra \lr |e\ra$ transition,
while the probe beam is detuned by an amount which is thrice the transition linewidth $\gamma_{ge}$. We do not
yet understand why this precise value shows a maximum nonlinearity. Perhaps the cause lies in the nature of
different coherent superpositions that are formed for different values of $\omega_p$ and $\omega_l$. 

The curves in figure 2, which are labeled 1 to 5, correspond to different values of
$\Omega_l=10^{-5},~1.0,~2.0,3.0$ and $4.0$. Parts (a) and (c) show Im$\chi^{(1)}$, while parts
(b) and (d) show corresponding variation in Re$\chi^{(1)}$. In all of them, $\Omega_p$ values
on the ordinate range from a $10^{-5}$ to 5.0, i.e., we investigate situations as
$\Omega_p$ ranges from a near-zero value, increasing upto becoming equal to $\Omega_l$ and then
increasing further. All values used for computation are normalized to the coherence decay
$\gamma_{ge}$. Note that this susceptibility is calculated to first order in probe beam ${\cal
E}_p$ but to all orders in pump beam amplitude ${\cal E}_l$.

Note that each curve begins with a different value of $\ch$, indicating that when $\Omega_p$ is nearly zero,
the value of $\ch$ depends only on $\Omega_l$.

 As $\Omega_p$ is increased, value of $\ch$ is determined by instantaneous values of both $\Omega_l$ and
 $\Omega_p$. For very large values of of $\Omega_p$, all curves asymptotically reach a single value of $\ch$,
irrespective of the $\Omega_l$ value. In other words, when the two Rabi frequencies differ by a large amount,
the larger of the two determine the behaviour of $\ch$. Varying one of them with the other one fixed will show
a behaviour of $\ch$ as seen in curves of figure 2, indicating the competition between two Rabi frequencies
which control the behaviour.

Curve number 1 of figure 2a corresponds to $\Omega_l=10^{-5}$, which is effectively a situation where only the
probe laser is present. Appropriately, this shows a saturation of probe absorption as $\Omega_p$ is increased. Curves 2 and
3, with $\Omega_l=1.0$ and $2.0$ respectively, show an increased absorption, which eventually saturates.
Surprisingly, increasing $\Omega_l$ still further leads to reduced absorption, as shown by curves 4 and 5 of
the same figure. The accompanying dispersion curves from figure 2 b show a corresponding behaviour. All the 5
curves asymptotically reach a saturation value of Re $\ch \approx 0$. But curve 1 shows an early saturation
behaviour as compared to others. Curves 2 and 3 begin with an increased dispersion and eventually reach the
saturation value, although at a much higher value of $\Omega_p$. Increasing $\Omega_l$ further, as shown by
Curves 4 and 5, dispersion flips the sign and becomes negative throughout, eventually approaching zero
asymptotically. This means that dispersion can be varied between negative and positive values by controlling
$\Omega_l$, and thus a propagating probe pulse can be compressed or expanded dynamically. Note that the
absolute magnitude of Re$\ch$ remains less than one, even when negative. Therefore the refractive index remains
positive for all values of $\Omega_p$. However, we can control the refractive index between values less than
one and greater than one, by choosing appropriate value of $\Omega_l$.

Plots of figure 2c and 2d show the data for same conditions except that a collisional decay rate of
$\gamma_c=0.1$ has been introduced. It is clearly seen that $\ch$ reaches its saturation value for a higher 
value of $\Omega_p$, than in case of no collisions. The numerical value of $\ch$ also significantly reduces,
than its counterpart of no collision case. The qualitative behaviour of the variation, including the changes in
sign of Re$\ch$, remains same. It has been shown that collisional processes destroy coherences formed and the
laser that creates the coherence is required to be stronger in order to balance this loss
\cite{arimondo_collision}. The effect seen here is similar, since it requires a stronger $\Omega_p$ to saturate
$\ch$ when collisions are present. This is an indication that the phenomena is indeed driven by atomic
coherence, rather than mere optical pumping.

 We now look at the 'Dressed states' to understand the behaviour seen in above graphs. We can safely assume that
whenever one of the two lasers are stronger than the other, it is the stronger laser which 'dresses' the energy
levels and the weaker laser probes the dressed atom. Therefore, when $\Omega_p << \Omega_l$, we look at the
atom as it is dressed by $\Omega_l$ laser. When $\Omega_p >> \Omega_l$, we look at the dressed states created
by $\Omega_p$ as probed by $\Omega_l$ laser. Complications arise only when both lasers are of comparable
intensity when dressed states formed by both lasers together has to be taken into account. Seen in this
viewpoint, the graphs in figure 2  indicate the behaviour of the system as it traverses between different
set of dressed states. In the beginning, when $\Omega_p$ is small, the system is in dressed states of
$\Omega_l$. As $\Omega_p$ is increased, the system goes through a complex set of dressed states formed due to
both $\Omega_p$ and $\Omega_l$ lasers and finally, reaches the dressed states of $\Omega_p$ laser. Intuitively,
one can understand that for small values of $\Omega_p$, different values of $\Omega_l$ form dressed eigenstates
with different energies and hence the probe absorption is different. When $\Omega_p$ reaches sufficiently high
value, the dressed states formed by $\Omega_p$ dominates and all atoms reach the CPT state formed by
$\Omega_p$. Thus the atoms are trapped and hence both absorption and dispersion reach a value of zero. 

In the following section, we look at the nature of the dressed states to substantiate this conclusion.

\section{Analysis in Dressed atom picture}
We consider three separate cases, viz, (i)$\Omega_l>>\Omega_p$, (ii)$\Omega_l\approx\Omega_p$ and
(iii)$\Omega_l<<\Omega_p$, and analyze the dressed states in each of them. 

\noindent
{\bf (i): $\Omega_l>>\Omega_p$: Dressed states due to $\Omega_l$:} We compute semiclassical dressed states
following \cite{cohentannoudji}. The complete Hamiltonian (\ref{hamiltonian}) is diagonalized and the
diagonalizing matrix is multiplied with the Bare state vector to obtain the dressed states as a superposition
of bare states. Further, this configuration can be divided into two subconfigurations, viz., a $\Lambda$ system
formed by levels $|g_-\rangle \lr |e_0\ra \lr |g_+\ra$ and a {\bf V} system formed by $|e_-\ra \lr |g_0\ra \lr
|e_+\ra$. Subsequently the eigensystems for both of them are given as 

\noindent for V($e_+,g_o,e_-$)
\begin{eqnarray}
\lambda^{V}_o &=& \Delta_l \nonumber \\
\lambda^{V}_{\pm} &=& \frac{\Delta_l \pm \beta_V}{2} \nonumber \\
\psi^V_o &=& \frac{1}{\sqrt{2}} ( |e_+\rangle - |e_-\rangle) \nonumber \\
\psi^V_{\pm} &=& \frac{(\Omega_+|e_+\rangle +\Omega_-|e_-\rangle \pm P_{\pm}|g_0\rangle}
{\sqrt{\alpha_{V}^2 + P_\pm^2}} \nonumber \\ 
{\rm where}&& \nonumber \\
\alpha_V^2 &=& |\Omega_+|^2 + |\Omega_-|^2 \nonumber\\
\beta_V &=& \sqrt{ \Delta_l^2 + 4 \alpha_{V}^2} \nonumber\\
{\rm and}~~ P_\pm&=& \frac{-\Delta_l \pm \beta_V}{2}.
\label{eigensystem1}
\end{eqnarray}

\noindent Similarly for $\Lambda$($g_+,e_o,g_-$)
\begin{eqnarray}
\lambda^{\Lambda}_o &=& 0 \nonumber \\
\lambda^{\Lambda}_{\pm} &=& \frac{\Delta_l \pm \beta_\Lambda}{2} \nonumber \\
\psi^\Lambda_o &=& \frac{1}{\sqrt{2}} ( |g_+\rangle - |g_-\rangle) \nonumber\\
\psi^\Lambda_{\pm} &=& \frac{\Omega_-^*|g_+\rangle - \Omega_+^*|g_-\rangle \pm Q_\pm |e_o
\rangle}{\sqrt{\alpha_{\Lambda}^2 + Q_\pm^2}} \nonumber \\
\alpha_\Lambda^2 &=& |\Omega_-|^2 + |\Omega_+|^2  \nonumber \\
 \beta_\Lambda &=& \sqrt{ \Delta_l^2
+ 4 \alpha_{\Lambda}^2} \nonumber\\
{\rm and}~~Q_\pm&=&\frac{\Delta_l \pm \beta_\Lambda}{2}.
\label{eigensystem2}
\end{eqnarray}

Also, $\Omega_+=\Omega_-=\Omega_l$. The levels $\psi^V_{0,\pm}$ and $\psi^{\Lambda}_{0,\pm}$ form a ladder of eigenstates, separated by a value
equal to $\omega_l$. There are nine possible coupling between these levels brought about by the probe laser,
but only four of them have nonzero transition probabilities. Quantum interferences causes cancellation of
transition in other five cases, as explained by (\ref{rules}) below

\begin{eqnarray}
\langle \psi^\Lambda_o | d.{\cal E}^o_p | \psi^V_o \rangle
;\mbox{cancellation}, \nonumber \\   
\langle \psi^\Lambda_o | d.{\cal E}^o_p | \psi^V_\pm \rangle
;\mbox{summation}, \nonumber \\ 
\langle \psi^\Lambda_\pm | d.{\cal E}^o_p | \psi^V_o \rangle
;\mbox{summation}, \nonumber \\ 
\langle \psi^\Lambda_\pm | d.{\cal E}^o_p | \psi^V_\pm \rangle
;\mbox{cancellation}.
\label{rules}
\end{eqnarray}

The level scheme, along with allowed transition is shown in figure 3. It is evident from
(\ref{eigensystem1},\ref{eigensystem2}), that the eigenvalues $\lambda^{\Lambda}_{0,\pm}$ and
$\lambda^{V}_{0,\pm}$, and hence the energy differences between them, depend upon value
$\Omega_{\pm}=\Omega_l$. In other words, different $\Omega_l$ values will induce different energy shifts
to the relevant energy levels and as a consequence, a probe beam coupling them will see different detunings.
This will produce different values of $\ch$ as seen in figure 2, in the regions of small $\Omega_p$.

\noindent 
{\bf (ii): $\Omega_l<<\Omega_p$: Dressed states due to $\Omega_p$:}
The situation of  $\Omega_p$ being stronger than $\Omega_l$ is similar to the one examined by Petrosyan and
coworkers \cite{petrosyan}\footnote{The authors of this paper investigate transient coupling between photons of
$\Omega_l$ and $\Omega_p$ through the atom and show that this coupling strongly depends upon the Zeeman shift
of $|g_\pm\ra$ and $|e_\pm\ra$ states by an external magnetic field. We present a steady state, zero-field
situation which is one of the limiting cases of our investigation}. In absence of $\Omega_l$ field, the dressed
state eigensystem is related to two two-level systems viz., $|g_\pm \rangle \leftrightarrow |e_\pm \ra$, given
by 
\begin{eqnarray}
|\Psi_{++}\ra&=&|e_+\ra +|g_+\ra \cr
|\Psi_{+-}\ra&=&|e_+\ra -|g_+\ra \cr
|\Psi_{-+}\ra&=&|e_-\ra +|g_-\ra \cr
|\Psi_{--}\ra&=&|e_-\ra -|g_-\ra 
\label{twolevel}
\end{eqnarray}

The $\Omega_p$ field does not couple $|g_0\ra \lr |e_0\ra$ transition. Atoms that reach $|g_0\ra$ due to
spontaneous emission from $|e_\pm\ra$ are therefore uncoupled from any radiation and are thus trapped in
$|g_0\ra$. However, such atoms are pumped out by $\Omega_l$ laser, which couples $|g_0\ra \lr |e_\pm\ra$
transitions. These transitions are shown by solid lines of figure 4.  As $\Omega_p$ is increased, the
process of optically pumping the atom into the trap state $|g_0\ra$ dominates over the process of
pumping it out by $\Omega_l$ laser, and hence the probe absorption eventually reaches zero, as seen in
figure 2.

\noindent
{\bf {iii}: $\Omega_l \approx \Omega_p$: Dressed states due to combined field:} In this condition it is hard to
obtain analytical expression for the dressed states since the two fields are detuned from each other. Only when
both fields are on resonance with the atomic transition and hence $\omega_l-\omega_p=0$, we can perform
rotational transformation which eliminates time dependency in Hamiltonian and hence analytically obtain
eigenvalues 

\begin{eqnarray}
\lambda_1&=&0.0,\\
\lambda_2 &=& \Delta_l,\\
\lambda_{3+} &=& \lambda_{4+} = \frac{\Delta_l + \sqrt{\Delta_l^2 +
2|\Omega_l|^2 + |\Omega_p|^2}}{2},\\ 
\lambda_{3-} &=& \lambda_{4-} = \frac{\Delta_l - \sqrt{\Delta_l^2 +
2|\Omega_l|^2 + |\Omega_p|^2}}{2}. 
\end{eqnarray}

We can show that there exists a three-component CPT state for this configuration, given by 
\begin{equation}
|\psi_1^o \rangle = \frac{(\frac{\Omega_p}{\Omega_l})|g_o \rangle - |g_- \rangle +
|g_+ \rangle }{\sqrt{ 2 + |\frac{\Omega_p}{\Omega_l}|^2}},
\label{eigen_threelevel}
\end{equation}
identical to one found in reference \cite{threecomponent}. The probability amplitude for $|g_0\ra$ is seen to
depend upon the ratio of Rabi frequencies of the two lasers $\Omega_p/\Omega_l$. On basis of this, it can be
interpreted that the occupation probability of $|g_0\ra$ state under these conditions depend upon the relative
strength of the two lasers. Intuitively extending this to the situation when $\omega_l-\omega_p \ne 0$, one can
assume that the superposition state deviates only little from (\ref{eigen_threelevel}). In other words, the
occupation probability of the state $|g_0\ra$ depends only on the relative intensities of the two lasers.

\noindent {\bf The mixed states} :
A study of Tr($\rho^2$) shows the distribution of atomic population among the different CPT states. Figure 5
shows Tr($\rho^2$) as a function of $\omega_l-\omega_p$, at different situations. The curves labeled 1 to 4
correspond to $\Omega_p=10^{-5},~1.0,~4.0$ and $6.0$. It can be noticed that Tr(${\rho^2}$) is close to 1.0
throughout when $\Omega_p=10^{-5}$, denoting that all the atomic population is populating one pure state. As
$\Omega_p$ is increased, the value of Tr(${\rho^2}$) is exactly equal to 1.0 only at large values of 
$|\omega_l-\omega_p|$,  or  when $\omega_l-\omega_p$ is exactly zero. For any values in between,
Tr(${\rho^2}$) deviates from 1.0 and the magnitude of this deviation depends upon value of $\Omega_p$. For all
the data shown in figure 5, $\omega_l$ is maintained at atomic resonance and $\omega_p$ is varied. The data
implies that the atomic system is in a pure state when $\omega_p$ is far away from $\omega_l$, which in other
words is the situation of a largely detuned $\Omega_p$. In such cases, the resonant $\Omega_l$ establishes a
stronger coupling and hence traps the atom in its CPT state. On the other hand,
when $\omega_p$ is exactly equal to $\omega_l$, and both beams are on resonance, the atom is in the three
component trap state (\ref{eigen_threelevel}). Tr(${\rho^2}$) being equal to 1.0 when
$\omega_l-\omega_p=0$, confirms this fact. For all values in between, Tr$({\rho^2})<1.0$ indicates that the
atom is in a mixed state, which means that the atomic population is distributed among many CPT states. 

In figure 6, we plot Im$\ch$ as a function of $\omega_l-\omega_p$. For each subset of graphs that are shown,
$\Omega_l$ is fixed and each curve corresponds to different $\Omega_p$, as described in figure caption. Since
the pump beam is kept at resonant, variation of $\omega_l-\omega_p$ is equivalent to changing the probe
detuning. It can immediately be seen that for small values of $\Omega_l$, the probe absorption is a
characteristic Lorentzian function of probe detuning, as shown in figure 6 (a). When the pump beam becomes
stronger, as in figures 6(b) and 6(c), the Rabi splitting of levels due to dynamic stark shift and onset of
Electromagnetically Induced Transparency is noticed by presence of side bands and zero absorption for
$\omega_l-\omega_p=0.0$.

It is now relevant to compare data shown in figure 6 with those in figure 2. The corresponding situation of
$\omega_l-\omega_p=3.0$ in figure 2 is equivalent to a point on the abscissa of figure 6. Increase of
$\Omega_p$, as in data of figure 2, is identical to a vertical line in figure 6 over which the point of
investigation moves. It can immediately be noticed that as $\Omega_p$ is increased, the peaks of Im $\ch$ curve
traverse across this vertical line. Since this peak shape and height also change across the curves of figure 6,
it is evident that the nonlinear Im $\ch$ seen in figure 2 originates in this traverse of peaks across the
vertical line.

In conclusion, we have studied an atomic response to a probe beam which varies nonlinearly with respect to the
strength of probe beam. The exact trajectory of this variation can be controlled using a control beam. The
control beam is polarized orthogonally to the probe beam and also propagates in a perpendicular direction. This
allows a very easy control of the two beams without affecting each other. The atomic system, with two sets of
degenerate triplets is often encountered in many alkali atoms. In short, this is a very simple and easy to use
system which allows a critical control and nonlinear variation of atomic response. Such critical control can
have applications in a wide ranging areas of information processing, magnetometry etc. A precise control on the
nonlinearity of $\ch$ will allow preparation and propagation of soliton pulses within the media as well.
Further, the phenomena 
also involves the atom being prepared in different CPT states. This property will be very useful in Quantum
Computation where a faithful preparation of a state is required before performing Computational operations on
it. Since these states are CPT states and constitute only ground states of the atom, such prepared states are
robust against intensity fluctuations and decoherence. Thus the state remains little affected till the Quantum
Computation operations are preformed. Preparations of such states also do not require precise
pulse shaping techniques. We shall follow up with a work describing such state preparation in detail. 

------------------

\centerline{\bf APPENDIX}
\def\ce{\cal E}
The 36-equations of motion of the
density matrix elements are obtained
using Liouville equation given in the
text along with the
relaxation  operator ${\cal L}\rho$ defined by
$$[{\cal L} \rho]_{lm}= \delta_{lm} \sum_{k \ne l} 2 \gamma_{lk}
\rho_{kk} -\left( \sum_{k \ne l} \gamma_{kl} + \sum_{k \ne m} \gamma_{km}
`\right ) \rho_{lm} \label{master},$$
where $l,m,k=g_+,g_0,g_-, e_+,e_0,e_-$.
We arrange the density matrix equation according to the 35 x 1 column
vector involving the elements of the density matrix $\rho$, in the form:
$$\matrix{ \rho=[\rho_{{g_-}{e_0}} & \rho_{{e_0}{g_+}} &
\rho_{{g_-}{g_+}}& \rho_{{g_-}{g_-}}& \rho_{{g_+}{g_+}}&
\rho_{{e_0}{g_-}}& \rho_{{e_0}{g_+}} & \rho_{{g_+}{g_-}}&
\rho_{{g_0}{e_-}}& \rho_{{e_+}{g_0}} \cr
\rho_{{e_-}{e_+}} & \rho_{{e_-}{e_-}} & \rho_{{e_+}{e_+}}
&\rho_{{e_0}{e_0}} & \rho_{{e_-}{g_0}}& \rho_{{g_0}{e_+}} &
\rho_{{e_-}{g_-}}& \rho_{{e_-}{g_+}}& \rho_{{e_-}{e_0}} &\rho_{{e_+}{g_+}}
\cr \rho_{{e_+}{g_-}} &\rho_{{e_0}{g_0}}& \rho_{{g_-}{g_0}}&
\rho_{{g_0}{g_+}}& \rho_{{g_-}{e_-}} & \rho_{{e_0}{e_-}} &
\rho_{{e_+}{g_+}}&\rho_{{g_-}{e_+}} & \rho_{{e_0}{e_+}}&
\rho_{{g_0}{e_0}} \cr
\rho_{{g_0}{g_-}}& \rho_{{g_+}{g_0}}]. }$$

Now as explained in the text, the transformation defined by
$\rho_{eg}\rightarrow \bar \rho_{eg} \exp(i \omega_l t),~ \rho_{ee}
\rightarrow \bar \rho_{ee}$ and $\rho_{gg} \rightarrow \bar \rho_{gg}$ is
made. Noting $\rho_{eg}=\rho_{ge}^\ast$, we give below the transformed
equation equivalent to (\ref{eq1}) of the text.

\begin{eqnarray*}
\partial_t \rho_{{g_-}{e_0}}&=&[-i-\gamma_{{g_-}{e_0}}]
\rho_{{g_-}{e_0}}-i\Omega_+(\rho_{{g_-}{g_-}}-\rho_{{e_0}{e_0}})+ i\Omega_p T_-\rho_{{e_-}{e_0}}-i\Omega_- \rho_{{g_-}{g_+}}\cr
\partial_t \rho_{{e_0}{g_+}} &=& [i\Delta_l-\gamma_{{g_+}{e_0}}]
\rho_{{e_0}{g_+}} + i\Omega_+ \rho_{{g_-}{g_+}} + i\Omega_-(
\rho_{{e_0}{e_0}}-\rho_{{g_+}{g_+}})- i\Omega_p T_+\rho_{{e_0}{e_+}} \cr
\partial_t \rho_{{g_-}{g_+}}&=&[i\Delta_{{g_-}{g_+}}-\gamma_{{g_-}{g_+}}]
\rho_{{g_-}{g_+}}+i\Omega_+\rho_{{e_0}{g_+}} + i\Omega_p T_-\rho_{{e_-}
{g_+}} - i \Omega_-\rho_{{g_-}{e_0}} - i \Omega_p T_+ \rho_{{g_-}{e_+}}\cr
\partial_t \rho_{{g_-}{g_-}}&=& -(2\gamma_{g_+g_-}+2 \gamma_{g_0g_-})
\rho_{{g_-}{g_-}}+ 2 \gamma_{{g_-}{e_-}}\rho_{{e_-}{e_-}}+2 \gamma_
{{g_-}{e_0}}\rho_{{e_0}{e_0}}+2 \gamma_{{g_0}{g_-}}\rho_{{g_0}{g_0}} \cr
&&+ 2 \gamma_{{g_-}{g_+}}\rho_{{g_+}{g_+}}   + i \Omega_+ \rho_{{e_0}
{g_-}}  + i \Omega_p T_-\rho_{{e_-}{g_-}} - i\Omega_+ \rho_{{g_-}{e_0}} - i \Omega_p T_+\rho_{{g_-}{e_-}} \cr
\partial_t \rho_{{g_+}{g_+}}&=& -(2\gamma_{g_-g_+}+2\gamma_{g_0g_+})
\rho_{{g_+}{g_+}}+ 2 \gamma_{{g_+}{e_+}}\rho_{{e_+}{e_+}} + 2
\gamma_{{g_+}{e_0}}\rho_{{e_0}{e_0}}+ 2 \gamma_{{g_-}{g_+}}
\rho_{{g_-}{g_-}} \cr 
&&+ 2 \gamma_{{g_0}{g_+}} \rho_{{g_0}{g_0}} 
+i \Omega_-  \rho_{{e_0}{g_+}} + i \Omega_p \rho_{{e_+}{g_+}} -
i \Omega_- \rho_{{g_+}{e_0}} - i
\Omega_p  \rho_{{g_+}{e_+}} \cr
\partial_t \rho_{{e_0}{g_-}}&=&[i\Delta_l-\gamma_{g_-e_0}]
\rho_{{e_0}{g_-}}+ i\Omega_+
(\rho_{{g_-}{g_-}}-\rho_{{e_0}{e_0}}) -
i\Omega_p \rho_{{e_0}{e_-}} + i \Omega_-  \rho_{{g_+}{g_-}} \cr
\partial_t \rho_{{g_0}{e_-}}&=& [i\Delta_l-\gamma_{{g_0}{e_-}}]\rho_{{g_0}
{e_-}}-i\Omega_-
(\rho_{{e_-}{e_-}}-\rho_{{g_0}{g_0}}) - i \Omega_p  \rho_
{{g_0}{g_-}}+ i\Omega_- \rho_{{e_+}{e_-}} \cr
\partial_t \rho_{{g_0}{e_+}}
&=&[i\Delta_l - \gamma_{g_0}{e_+}]\rho_{{g_0}{e_+}} -i \Omega_-
(\rho_{{g_0}
{g_0}}-\rho_{{e_+}{e_+}}) + i\Omega_+ \rho_{{e_-}{e_+}} - i\Omega_p \rho_
{{g_0}{g_+}} \cr
\partial_t \rho_{{e_-}{e_+}}&=&-\gamma_{{e_-}{e_+}}\rho_
{{e_-}{e_+}}+i\Omega_p \rho_{{g_-}{e_+}}+i
\Omega_- \rho_{{g_0}{e_+}}-i
\Omega_p \rho_{{e_-}{g_+}} - i\Omega_+ \rho_{{e_-}{g_0}}\cr
\partial_t \rho_{{e_-}{e_-}}&=&-(2\gamma_{g_0e_-}+2\gamma_{g_-e_-})\rho_{{e_-}{e_-}} +2\gamma_{{e_-}{e_0}}
\rho_{{e_0}{e_0}}+
2\gamma_{{e_-}{e_+}} \rho_{{e_+}{e_+}}+ i \Omega_p \rho_
{{g_-}{e_-}} \cr
&& + i \Omega_p \rho_{{e_-}{g_-}} - i \Omega_- \rho_{{e_-}{g_0}} \cr
\partial_t \rho_{{e_+}{e_+}} &=& -(2\gamma_{g_+e_+}+2\gamma_{g_0e_+})\rho_{{e_+}{e_+}}+2
\gamma_{{e_-}{e_+}}\rho_{{e_-}{e_-}} + 2
\gamma_{{e_+}{e_0}} \rho_{{e_0}{e_0}} + i \Omega_p \rho_{{g_+}{e_+}} \cr
&&+ i \Omega_+ \rho_{{g_0}{e_+}} - i\Omega_p \rho_{{e_+}{g_+}} - i \Omega_+ \rho_{{e_+}{g_0}}\cr
\partial_t \rho_{{g_0}{g_0}}&=&-(2\gamma_{g_+g_0}+2\gamma_{g_-g_0}) \rho_{{g_0} {g_0}} + 2
\gamma_{{g_-}{g_0}} \rho_{{g_0}{g_0}} + 2 \gamma_{{g_+}{g_0}} \rho_{{g_+}{g_+}} + 2
\gamma_{{g_0}{e_-}}\rho_{{e_-}{e_-}} \cr &&
+ 2\gamma_{{g_0}{e_+}} \rho_{{e_+}{e_+}} + i \Omega_- \rho_{{e_-}{g_0}} + i \Omega_+ \rho_{{e_+}{g_0}}
    - i \Omega_- \rho_{{g_0}{e_-}} -i \Omega_+ \rho_{{g_0}{e_+}} \cr
\partial_t \rho_{{e_-}{g_-}}&=&[i\Delta_l
- \gamma_{{e_-}{g_-}}]\rho_{{e_-}{g_-}} - i \Omega_p
(\rho_{{e_-}{e_-}}-
\rho_{{g_-}{g_-}}) + i \Omega_- \rho_{{g_0}{g_-}} - i \Omega_+ \rho_{{e_-}
{e_0}}\cr
\partial_t \rho_{{e_-}{g_+}}&=& [i\Delta_l-\gamma_{{e_-}{g_+}}]
\rho_{{e_-}{g_+}} +i \Omega_p
\rho_{{g_-}{g_+}} + i \Omega_- \rho_{{g_0}{g_+}}
 - i\Omega_- \rho_{{e_-}{e_0}} - i\Omega_p \rho_{{e_+}{e_-}}\cr
\partial_t \rho_{{e_-}{e_0}}&=& [i\Delta_l-\gamma_{{e_-}{e_0}}]\rho_{{e_-}{e_0}} +i\Omega_p \
\rho_{{g_-}{e_0}} + i \Omega_-
\rho_{{g_0}{e_0}} - i\Omega_+ \rho_{{e_-}
{g_-}} - i \Omega_- \rho_{{e_-}{g_+}}\cr
\partial_t \rho_{{e_+}{g_+}}&=&
[i\Delta_l-\gamma_{{e_+}{g_+}}]\rho_{{e_+}{g_+}}-i\Omega_p
(\rho_{{e_+}
{e_+}}-\rho_{{g_+}{g_+}}) + i\Omega_+ \rho_{{g_0}{g_+}} - i\Omega_-\rho_
{{e_+}{e_0}}\cr
\partial_t \rho_{{e_+}{g_-}} &=&[-i\Delta_l-(\gamma_{e_+}
+\gamma_{g_+})]\rho_{{e_+}{g_+}}-i
\Omega_+ \rho_{{e_+}{e_0}} + i \Omega_-
\rho_{{g_0}{g_-}} + i \Omega_p \rho_{{g_+}{g_-}} - i
\Omega_p \rho_{{e_+}
{e_-}}\cr
\partial_t \rho_{{e_+}{e_0}}&=& [-i\Delta_l-\gamma_{{e_+}{e_0}}]
\rho_{{e_+}{e_0}}-i \Omega_+
\rho_{{e_+}{g_-}} - i\Omega_- \rho_{{e_+}{g_+}}
+ i \Omega_+ \rho_{{g_0}{e_0}} + i \Omega_{{e_+}{g_+}}
\rho_{{g_+}{e_0}}\cr
\partial_t \rho_{{e_0}{g_0}}&=&[-i\Delta_l -\gamma_{{e_0}{g_0}}]\rho_{{e_0}
{g_0}} +i \Omega_+
\rho_{{g_-}{g_0}} + i \Omega_- \rho_{{g_+}{g_0}} - i
\Omega_- \rho_{{e_0}{e_-}} - i \Omega_+
\rho_{{e_0}{e_+}}\cr
\partial_t
\rho_{{g_-}{g_0}}&=& [i\Delta_l-\gamma_{{g_-}{g_0}}]\rho_{{g_-}{g_0}}+ i
\Omega_+
\rho_{{e_0}{g_0}} + i \Omega_p \rho_{{e_-}{g_0}} - i \Omega_-
\rho_{{g_-}{e_-}} - i \Omega_+
\rho_{{g_-}{e_+}}\cr
\partial_t
\rho_{{g_+}{g_0}}&=&[i\Delta_l-\gamma_{{g_+}{g_0}}]\rho_{{g_+}{g_0}}+
i \Omega_-
\rho_{{e_0}{g_0}} + i \Omega_p \rho_{{e_+}{g_0}} - i
\Omega_+ \rho_{{g_+}{e_+}} - i \Omega_-
\rho_{{g_+}{e_-}}\cr
&&{\rm and ~rest ~of ~the ~equations ~are ~obtained ~by,}\cr
\rho_{mn}&=&\rho_{nm}^\ast
\end{eqnarray*}

Where, $T_\pm=\pm\exp(i \omega_l t)$. $\gamma_{ij}$ is the coherence dephasing
term of $\rho_{ij}$. Detuning is defined as
$\Delta_l=\omega_e-\omega_g-\omega_l$,  where $\omega_e$ and $\omega_g$ are
energy of levels $|e\rangle$ and $|g\rangle$ in angular units.
$\rho_{{g_0}{g_0}}$ is eliminated using  equation (\ref{constr}) of the text.
Then the terms containing $T_+$ form part of the  35 x 35 matrix $M_+$.Those
containing $T_-$ form part of 35 x 35 matrix  $M_-$. Terms independent of these
form the 35 x 35 matrix $A$. The nonzero elements of $I_0$, however, are given
by
\begin{eqnarray*}
I_0(1) =i\Omega_{{e_0}{g_-}}^\ast&;& I_0(2) = -i\Omega_{{e_0}{g_+}} \cr
I_0(4) =I_0(5) = \gamma_{{e_0}{g_-}} &;& I_0({g_+}) = -i\Omega_{{e_0}{g_-}}\cr
I_0(7) = i\Omega_{{e_0}{g_+}}^\ast &;& I_0(12) = \gamma_{{e_0}{e_-}} \cr
I_0(13) &=& \gamma_{{e_0}{e_+}} \cr
\end{eqnarray*}

Then the matrices $A,~M_+,~M_-$ and $I_0$ are obtained accordingly. $A,
~M_+$ and $M_-$ are 35 x 35 matrices each and  $I-0$ is a 35 x 1 matrix in
the same order. The nonzero elements of $A,~M_+,~M_-$ are too many in number
to be elaborated here.

\newpage

{\bf Figure Captions}

\begin{itemize}
\item[Fig. 1] The six level energy level model used in the text.
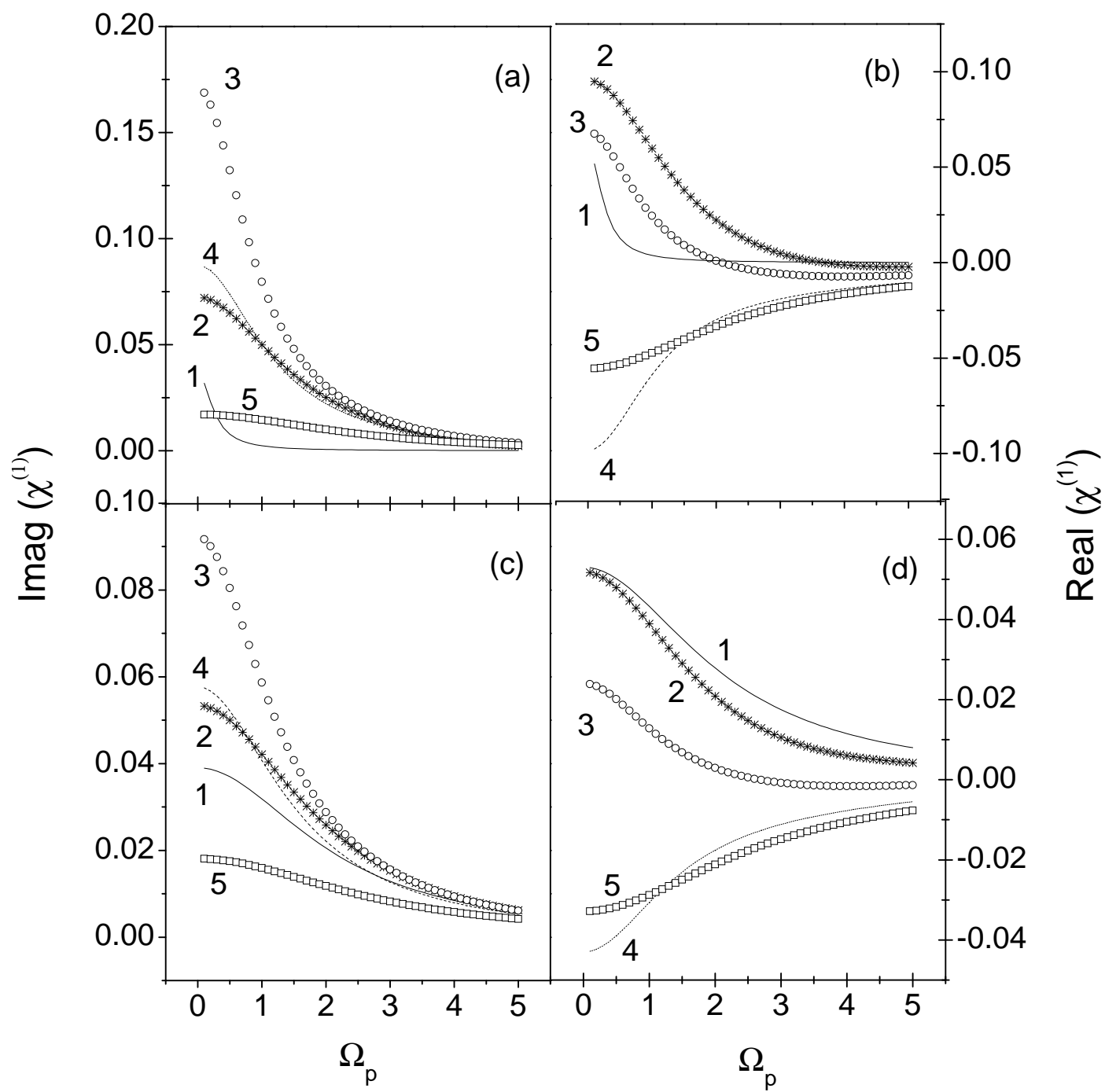
\item[Fig. 2] Real and imaginary $\chi^{(1)}$ as a function of $\Omega_p$, for
different values of $\Omega_l$. (a) and (c) correspond to absorption and (b) and
(d) correspond to dispersion determined from equation (14) of the text.
The labeled curves are for the $\Omega_l$ values (1)$10^{-5}$, (2) 1.0, (3) 2.0
(4) 3.0 (5) 4.0. $\Delta_l=0.0$ and $\omega_l-\omega_p=3.25$.
$\gamma_c=10^{(-5)}$ for curves in (a) and (b) and $\gamma_c=0.1$ for curves in
(c) and (d). All numerical values are normalised to $\gamma_{eg}$.

\item[Fig. 3] Bare states (- - - lines), dressed states formed due to
$\Omega_l$ ( --- lines) and allowed transitions between them
(represented by oblique $\longrightarrow$).
\item[Fig. 4] Bare states (- - -  lines), dressed states formed
due to strong $\omega_p$ beam (--- lines) and allowed transitions between
them brought about by the $\omega_l$ beam (oblique $\longrightarrow$).
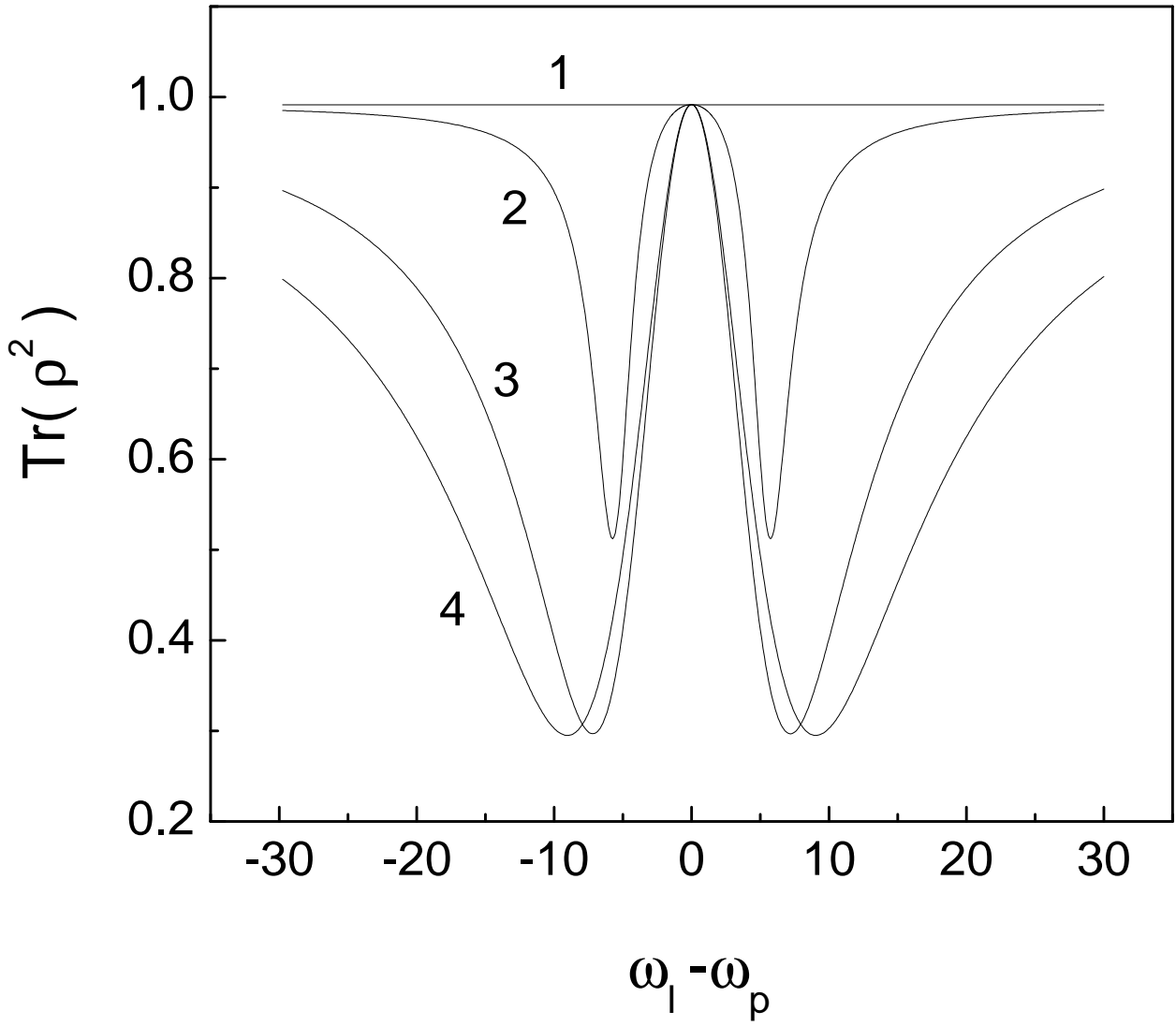
\item[Fig. 5] Time averaged Tr$(\rho^2)$ for different values of Rabi
frequencies as a function of relative detuning $(\omega_l-\omega_p)$.
The labeled curves correspond to different $\Omega_p$ values:
(1)$10^{-5}$,(2)1.0, (3)4.0 and (4)6.0.
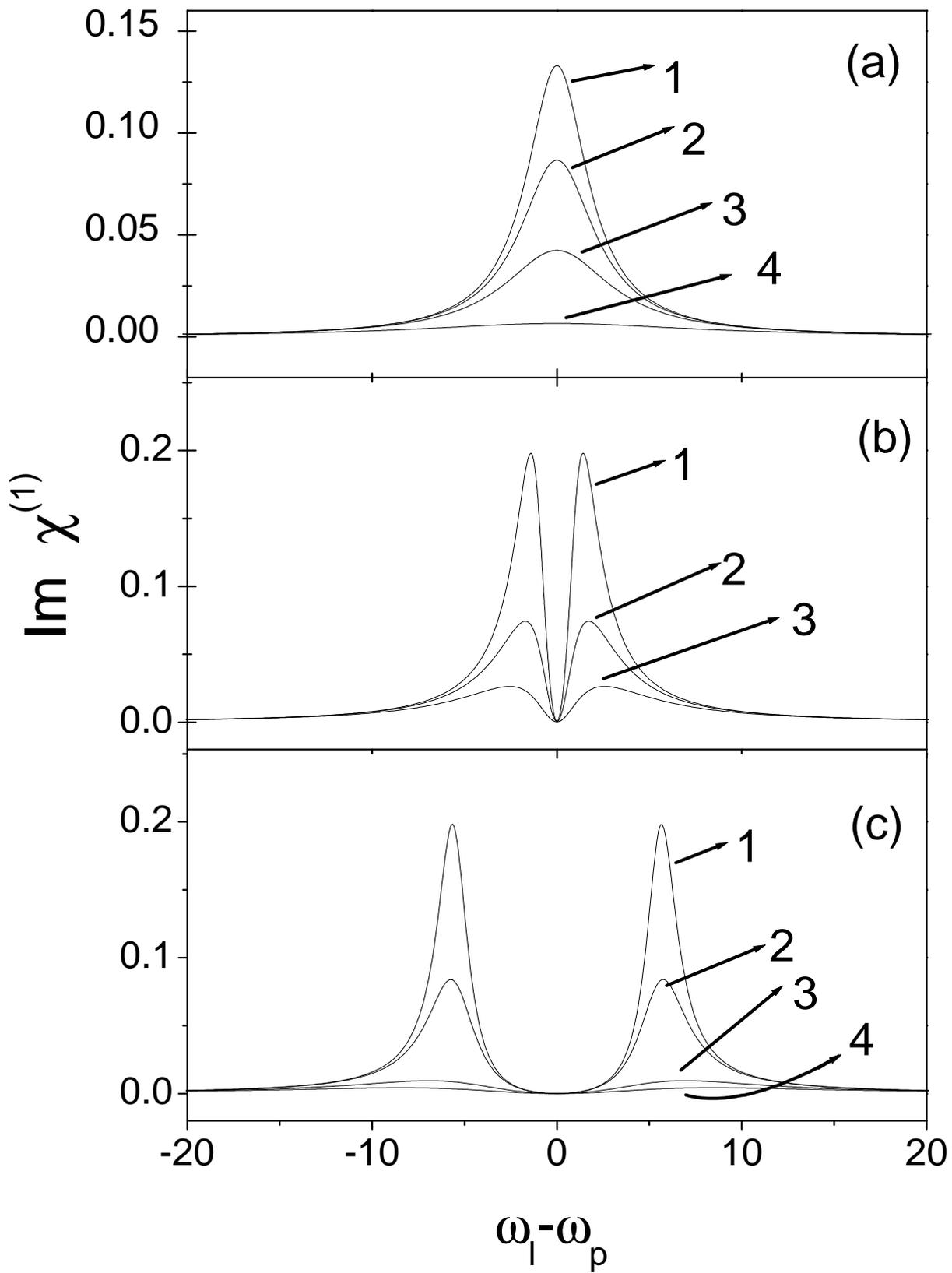
\item[Fig. 6]Absorption of probe beam as a function of relative detuning
$(\omega_l-\omega_p)$ for a fixed $\Omega_l$ value and different values of
$\Omega_p$. For figure (a), $\Omega_l=10^{-5}$ and the labelled curves are
for values of $\Omega_p$ values equal to (1) $10^{-5}$, (2) 1.0, (3) 0.2
and (4) 0.6. For figure (b), $\Omega_l=1.0$ and the labelled curves
are for $\Omega_p=$ (1) $10^{-5}$, (2) 1.0 and (3)2.0. For figure (c),
$\Omega_l=4.0$ and the labelled curves are for $\Omega_p$ values
(1)10$^{-5}$, (2) 1.0, (3) 4.0 and (4) 6.0. The transparency is
observed only when $\Omega_l$ is strong. $\gamma_c=10^{-3}$ for all the
three cases. 
\end{itemize}

\begin{thebibliography}{00}
\bibitem{harris}S. E. Harris and Lene Vestergaard Hau 1999 {\em Phys. Rev. Lett}
{\bf 82} 4611
\bibitem{petrosyan}D. Petrosyan and G. Kurizki 2002 {\em Phys. Rev. }{\bf A 65} 33833 
\bibitem{kerr}H. Schmidt and A. Imamo\u glu 1996 {\em Opt. Lett.} {\bf 21} 1936 
\bibitem{spt} Agarwal G. S. and Surya P. Tewari 1986 {\em Phys. Rev. Lett.} {\bf 56} 1811\\
Agarwal G. S. and Surya P. Tewari 1991 {\em Phys. Rev. Lett.} {\bf 66} 1797 \\
Agarwal G. S. and Surya P. Tewari 1993 {\em Phys. Rev. Lett.} {\bf 70} 1417
\bibitem{maneesh} M. Jain, A. J. Meriam, A. Kasapi, G. Y. Yin and S. E. Harris	1995 {\em Phys. Rev. Lett.} {\bf 75}
4385
\bibitem{dunn}R. R. Moseley,S.Shepherd, D. J. Fulton, B. D.Sinclair, Malcom H. Dun 1995 {\em Phys. Rev. Lett.} {\bf 74}
670 
\bibitem{switch}Min Xiao, Hai Wang, and David Goorskey 2002 {\em Optics and Photonics news} {\bf 13} 45 \\
Hai Wang, David Goorskey and Min Xiao 1002 {\em Opt. Lett.} {\bf 27}, 1354 \\
Andrew M. C. Dawees, Lucas Illing, Susan Clarck and Daniel Gauthier 2005 {\em Science} {\bf 308} 672 
\bibitem{strekalov}Dmitry Strekalov, Andrey B. Matsko and Lute Maleki 2005 {\em J. Opt. Soc. Am} {\bf B 22} 65
\bibitem{arimondo} an exhaustive review of CPT states can be found in "Coherent Population Trapping in Laser
Spectroscopy" E. Arimondo in {\em Progress in Optics
vol. XXXV}, ed: E. Wolf, Elsevier 1996 pp 259.
\bibitem{risken}H.Risken {\em Fokker Plank Equation},  Springer Verlag, Berlin (1984), pp 196. 
\bibitem{arimondo_collision}Arimondo. E. {\em Fundamentals of Quantum Optics
III}, ed: F. Ehlotzky (Lecture notes in Physics vol. 420), Springer Verlag,
Berlin, 1993, pp170.
\bibitem{cohentannoudji}C. Cohen-Tannoudji and S. Reynaud 1977 {\em J. Phys} {\bf B 10} 345
\bibitem{threecomponent}K. K. Meduri, G. A. Wilson, P. B. Sellin and T. W.
Mossberg 1993 {\em Phys. Rev. Lett.} {\bf 71} 4311 
\bibitem{boller} K. -J. Boller, A. Imamo\u glu and S. E. Harris 1991 {\em Phys. Rev. Lett.} {\bf 66}, 2593.
\end{thebibliography}
\end{document}